\documentclass{jfm}

\usepackage{graphicx}
\usepackage{amssymb}
\usepackage{amsmath}
\usepackage{pxfonts}
\usepackage{tikz,pgfplots}
\usepackage{scalerel} 
\usepackage{subcaption} 
\usepackage{lineno} 
\usepackage{verbatim} 
\usepackage[normalem]{ulem} 
\usetikzlibrary{shapes,snakes}

\newcommand{\dd}{\mathrm{d}}
\newcommand{\Eb}{\mathbb{E}}
\newcommand{\Cb}{\mathbb{C}}
\newcommand{\Sb}{\mathbb{S}}
\newcommand{\Ib}{\mathbb{I}}
\newcommand{\SSc}{\mathcal{S}}
\newcommand{\C}{\mathcal{C}}
\newcommand{\dx}{\mathrm{d}x}
\newcommand{\dy}{\mathrm{d}y}

\newcommand{\ou}{\overline{u}}
\newcommand{\ov}{\overline{v}}

\begin{document}

\shorttitle{Integral relations for the skin-friction coefficient}
\title{Integral relations for the skin-friction \\ coefficient of canonical flows}

\author{Pierre Ricco$^1$\corresp{\email{p.ricco@sheffield.ac.uk}} and Martin Skote$^2$}

\affiliation{$^1$Department of Mechanical Engineering, The University of Sheffield, \\ Sheffield, S1 3JD, United Kingdom \\ $^2$School of Aerospace, Transport and Manufacturing, \\ Cranfield University, Cranfield, MK43 0AL, United Kingdom}
\maketitle

\begin{abstract}
We show that the \cite{fukagata-iwamoto-kasagi-2002}'s identity for free-stream boundary layers simplifies to the von K\'arm\'an momentum integral equation relating the skin-friction coefficient and the momentum thickness when the upper bound in the integrals used to obtain the identity is taken to be asymptotically large. If a finite upper bound is used, the terms of the identity depend spuriously on the bound itself. Differently from channel and pipe flows, the impact of the Reynolds stresses on the wall-shear stress cannot be quantified in the case of free-stream boundary layers because the Reynolds stresses disappear from the identity. The infinite number of alternative identities obtained by performing additional integrations on the streamwise momentum equation also all simplify to the von K\'arm\'an equation. Analogous identities are found for channel flows, where the relative influence of the physical terms on the wall-shear stress depends on the number of successive integrations, demonstrating that the laminar and turbulent contributions to the skin-friction coefficient are only distinguished in the original identity discovered by \cite{fukagata-iwamoto-kasagi-2002}. In the limit of large number of integrations, these identities degenerate to the definition of skin-friction coefficient and a novel twofold-integration identity is found for channel and pipe flows. In addition, we decompose the skin-friction coefficient uniquely as the sum of the change of integral thicknesses with the streamwise direction, following the study of \cite{renard2016theoretical}. We utilize an energy thickness and an inertia thickness, which is composed of a thickness related to the mean-flow wall-normal convection and a thickness linked to the streamwise inhomogeneity of the mean streamwise velocity. The contributions of the different terms of the streamwise momentum equation to the friction drag are thus quantified by these integral thicknesses.
\end{abstract}


\section{Introduction}

Wall-bounded turbulent flows play a crucial role in an immense range of technological and industrial fluid systems, e.g. over vehicles moving in air and water, through pipes and ducts used for oil and gas transport and inside combustion and jet engines. Free-stream turbulent boundary layers are particularly relevant in aerodynamics applications and, with respect to flat-wall channel flows and circular pipe flows, they present an additional difficulty because the streamwise direction is statistically inhomogeneous as the shear-layer thickness grows downstream. A major research objective is the accurate computation or measurement of the streamwise evolving wall-shear stress in turbulent boundary layers. This task is more challenging than in pressure-driven channel and pipe flows because the wall friction cannot be determined conveniently through the streamwise pressure gradient, but only through the mean-velocity gradient at the wall.

A breakthrough in this research area has been the discovery of the Fukagata-Iwamoto-Kasagi identity (FIK) \citep{fukagata-iwamoto-kasagi-2002}, which relates the wall-shear stress to a simple integral involving the Reynolds stresses in channel and pipe flows, with the addition of other integral terms in the case of free-stream boundary layers because of their streamwise inhomogeneity. The integrals in the FIK identity are performed along the wall-normal direction from the wall to an upper integration bound, i.e. the flow centreline for channels and pipe flows, and the boundary-layer thickness for boundary layers. Another relevant identity was discovered by \cite{renard2016theoretical} (hereafter referred to as the RD decomposition), for which the skin-friction coefficient is expressed as the sum of integral terms belonging to the mechanical energy equation. Alternative identities for the skin-friction coefficients of these flows, derived from the vorticity equation, was obtained by \cite{yoon-etal-2016}, and variants for open-channel flows were studied by \cite{nikora-etal-2019} and \cite{duan-etal-2021}.

The utilization of the FIK decomposition for turbulent channel flows has been significant. It has also been used in the context of drag reduction techniques, for which it is important to understand the contribution of various quantities to the skin friction. It appeared in the studies of \cite{xia-etal-2015} and \cite{stroh-etal-2015} on boundary layers with opposition control, \cite{kametani-fukagata-2011}, \cite{kametani-etal-2015} and \cite{kametani-etal-2016}, where blowing and suction were used as the control mechanism, and \cite{bannier-etal-2015}, who analyzed flows with drag reduction by riblets. The influence of the large scale structures in the boundary layer was investigated with the aid of FIK decomposition by \cite{deck_renard_laraufie_weiss_2014}. \cite{monte-etal-2011} studied the flow over a cylinder to investigate the influence of the curvature ratio on the skin friction using the FIK identity.

The RD decomposition has recently become more popular in the study of boundary-layer flows. \cite{fan-li-pirozzoli-2019} used it to investigate incompressible and compressible turbulent boundary layers, focussing on the Reynolds-number behaviour of the different terms of the decomposition. \cite{fan-etal-2020} utilized the RD decomposition to study an adverse-pressure-gradient boundary layer while \cite{fan_atzori_vinuesa_gatti_schlatter_li_2022} investigated the flow over the suction and pressure sides of an airfoil. \cite{zhang-etal-2020} compared the application of the FIK and RD decompositions in channel flows with drag reduction due to viscoelastic fluids.

The interesting study by \cite{elnahhas-johnson-2022} is particularly worth mentioning because their identity expresses the skin-friction coefficient of free-stream boundary layers as the sum of the Blasius friction coefficient and an integrated contribution of the Reynolds stresses, thereby distinguishing the contribution of the laminar flow and the nonlinear fluctuations in transitional or turbulent boundary layers.

The choice of the boundary-layer thickness as the upper integration bound in the FIK analysis in the case of free-stream boundary layers was questioned by \cite{renard2016theoretical} because the definition of the thickness is arbitrary and the contribution of the turbulent fluctuations above that wall-normal location, albeit small, is thus neglected without justification. The impact of the upper integration limit on the terms of the identity was discussed by \cite{mehdi-etal-2014} and \cite{wenzel-etal-2022}. 

We show herein that, in the case of free-stream boundary layers, a finite upper bound of integration in the free stream generates a spurious dependence of the terms of the FIK identity on the bound itself. It follows that the upper bound has to be taken asymptotically large, a step that simplifies the FIK identity to the well-known von K\'arm\'an momentum integral equation relating the wall-shear stress and the momentum thickness. The influence of the Reynolds stresses on the wall friction cannot thus be quantified, as in the cases of channel and pipe flows. We also find that the infinite number of identities obtained by successive integration all reduce to the von K\'arm\'an momentum equation for boundary layers, while, for channel flows, only the original FIK identity possesses a clear physical meaning. By asymptotic analysis, it is revealed that the family of identities for channel flows collapses to the definition of skin-friction coefficient when the number of iterations increases to infinity. We interpret the skin-friction coefficient decomposition for boundary layers by \cite{renard2016theoretical} in terms of integral thicknesses, by utilizing an energy thickness and an inertia thickness, the latter composed of two thicknesses related to the mean-flow wall-normal convection and the streamwise inhomogeneity.

\section{Flow systems}

We consider a free-stream boundary layer flowing past a flat plate in the absence of a streamwise pressure gradient. Unless otherwise stated, the Navier-Stokes equations are scaled by using the free-stream velocity $U_\infty^*$ as the reference velocity and $\nu^*/U_\infty^*$ as the reference length scale, where $\nu^*$ is the kinematic viscosity of the fluid. Quantities denoted by $*$ are dimensional, while quantities without any symbol are non-dimensional. The Cartesian coordinates $x$, $y$, $z$ denote the streamwise, wall-normal and spanwise directions, respectively. The velocity components along $x$, $y$ and $z$ are $u$, $v$ and $w$, respectively. 
The flat plate is at $y=0$ and the flow is unconfined along the wall-normal direction. It is assumed that the flow has reached fully developed conditions and the direction $z$ and the time $t$ are statistically homogeneous. Averaging a quantity $q$ over $z$ along a distance $L_z$ and over $t$ for a time interval $T$ is defined as $\overline{q}(x,y) = L_z^{-1}T^{-1}\int_0^T \int_0^{L_z} q(x,y,z,t) \mathrm{d}z \mathrm{d}t$. Each quantity is decomposed as $q(x,y,z,t) = \overline{q}(x,y) + q'(x,y,z,t)$ and $\{\overline{u},\overline{v},0\}$ is the mean flow. The data obtained by \cite{sillero2013one} via direct numerical simulations are used. 
We also study integral relations for channel flows by using the data computed by \cite{hoyas-jimenez-2006} via direct numerical simulations.

\section{Results}

\subsection{Derivation of the momentum-thickness law}

It is first useful to review the derivation of the relationship between the skin-friction coefficient and the momentum thickness for free-stream boundary layers. The Reynolds-averaged $x$-momentum equation is
\begin{equation}
\label{eq:rans-x}
\frac{\p}{\p y}\left( \overline{u'v'} - \frac{\p \overline{u}}{\p y} \right) + I_x = 0, 
\quad \mbox{where} \quad 
I_x(x,y) = \frac{\p \overline{u u}}{\p x} + \frac{\p \overline{u} \ \overline{v}}{\p y} - \frac{\p^2 \overline{u}}{\p x^2}.
\end{equation}
Integrating \eqref{eq:rans-x} along $y$ from 0 to $\infty$ leads to
\begin{equation}
\label{eq:fik}
\frac{\p \overline{u}}{\p y}\biggm|_{y=0} = - \int_0^\infty I_x \dy = - \int_0^\infty \frac{\p }{\p x} \left( \overline{u u} - \frac{\p \overline{u}}{\p x} \right) \dy
\end{equation}
because $\overline{u'v'}$$\rightarrow$$0$, $\overline{v}$$\rightarrow$$0$ and $\overline{u}$$\rightarrow$$1$ as $y$$\rightarrow$$\infty$. The limit of vanishing free-stream wall-normal velocity is discussed in Appendix \ref{app:a}.
In this section and in \S\ref{sec:rd}, it is assumed that $\p \overline{u' u'}/\p x \ll \p \overline{u' v'}/\p y$ because in a turbulent boundary layer the correlations $\overline{u' u'}$ and $\overline{u' v'}$ are both comparable to the square of the wall-friction velocity $u_\tau^2=(\nu^*/U_\infty^{*2})\mathrm{d} \overline{u}^* /\mathrm{d}y^*$ and the derivative with respect to $x$ is negligible relative to the derivatives with respect to $y$ in the limit of large Reynolds number. This assumption has been amply verified numerically ever since the first direct numerical simulation of a spatially developing boundary layer by \cite{spalart-watmuff-1993}. By using the continuity equation, it follows that
\begin{equation}
\frac{\p \overline{u}}{\p y}\biggm|_{y=0} 
= - \int_0^\infty \frac{\p }{\p x} \left( \overline{u u} + \frac{\p \overline{v}}{\p y} \right) \dy
= - \int_0^\infty \left( \frac{\p \overline{u} \ \overline{u}}{\p x} +\frac{\p \overline{u' u'}}{\p x} \right) \dy
= - \int_0^\infty \frac{\p \overline{u} \ \overline{u}}{\p x} \dy.
\end{equation}    
By using the definition of momentum thickness
\begin{equation}
\label{eq:momentum-thickness}
\theta = \int_0^\infty  \overline{u}(1-\overline{u}) \dy,
\end{equation}
one finds
\begin{equation}
\label{eq:theta-pierre}
\frac{\p \overline{u}}{\p y}\biggm|_{y=0} 
= \frac{\mathrm{d}}{\dx} \int_0^\infty \overline{u}(1-\overline{u}) \dy
= \frac{\mathrm{d} \theta}{\dx}.
\end{equation}
Equation \eqref{eq:theta-pierre} can be written in terms of the skin-friction coefficient, 
\begin{equation}
\label{eq:theta-2}
C_f=\frac{2\nu^*}{U_\infty^{*2}}\frac{\p \overline{u}^*}{\p y^*}\biggm|_{y^*=0} 
=2\frac{\mathrm{d} \theta^*}{\mathrm{d}x^*},
\end{equation}
more commonly referred to in the literature as the von K\'{a}rm\'{a}n momentum integral equation \citep{pope-2000}. Further details of the derivation are found in Appendix \ref{app:a}. By integrating \eqref{eq:theta-2} along $x^*$, one finds
\begin{equation}
\label{eq:drag}
\mathcal{D}^* 
= 
\mu^* \int_{x_1^*}^{x_2^*} \frac{\p \overline{u}^*}{\p y^*}\biggm|_{y^*=0} \mathrm{d}x^* 
= 
\rho^* U_\infty^{*2} \left(\theta^*_2 - \theta^*_1\right),
\end{equation}
where $\mathcal{D}^*$ is the drag per unit spanwise width along a streamwise interval $x_2^*-x_1^*$, $\mu^*$ is the dynamic viscosity of the fluid and $\rho^*$ is the density of the fluid.

\subsection{Simplification of the FIK identity}
\label{sec:fik}

We rederive the FIK identify for a free-stream boundary layer following \cite{fukagata-iwamoto-kasagi-2002} with two important differences. The first difference is that \cite{fukagata-iwamoto-kasagi-2002} scaled $y^*$ by the boundary-layer thickness $\delta_{99}^*$, i.e. the wall-normal distance where the streamwise mean velocity $\overline{u^*}$ reaches 99\% of the free-stream velocity $U_\infty^*$, while we scale $y^*$ with $\nu^*/U_\infty^*$. The second difference is that \cite{fukagata-iwamoto-kasagi-2002} performed integration along $y$ from the wall to $\delta_{99}$, while we integrate from the wall to an unspecified location $h$ in the free stream and then take the limit $h$$\rightarrow$$\infty$.

Integrating \eqref{eq:rans-x} from 0 to $y$ leads to
\begin{equation}
\label{eq:fik-1}
\overline{u'v'} - \frac{\p \overline{u}}{\p y} + \frac{\p \overline{u}}{\p y}\biggm|_{y=0} + \int_0^y I_x \mathrm{d}{\hat y} = 0.
\end{equation}
By further integrating \eqref{eq:fik-1} from 0 to $y$, one finds
\begin{equation}
\label{eq:fik-2}
y \frac{\p \overline{u}}{\p y}\biggm|_{y=0} = - \int_0^y \overline{u'v'} \mathrm{d}\hat y + \overline{u} 
- \int_0^y \int_0^{\tilde y} I_x \mathrm{d}\hat y \mathrm{d} \tilde y.
\end{equation}
Integration of \eqref{eq:fik-2} from 0 to $h$, where $\overline{u} =1$ and $\overline{v}=0$, gives
\begin{equation}
\label{eq:fik-3}
\frac{h^2}{2}\frac{\p \overline{u}}{\p y}\biggm|_{y=0} = 
- \int_0^h \int_0^y \overline{u'v'} \mathrm{d}\hat y \dy 
+ \int_0^h \overline{u} \dy 
- \int_0^h \int_0^y \int_0^{\tilde y} I_x \mathrm{d}\hat y \mathrm{d} \tilde y \dy,
\end{equation}
and, by integrating by parts the first and the last term on the right-hand side of \eqref{eq:fik-3}, one finds
\begin{equation}
\label{eq:int-1-2}
C_f = 
\underbrace{\frac{4}{h^2} \int_0^h (y-h) \overline{u'v'} \dy}_{\mbox{term 1}}+ 
\underbrace{\frac{4}{h^2} \int_0^h \overline{u} \dy}_{\mbox{term 2}}- 
\frac{2}{h^2} \int_0^h (y-h)^2 I_x \dy.
\end{equation}
Equation \eqref{eq:int-1-2} coincides with the steady version of equation (15) in \cite{fukagata-iwamoto-kasagi-2002} if the wall-normal distance is scaled as $y_{99}=y^*/\delta_{99}^*$ and the upper bound $h$ is set equal to $\delta_{99}$, i.e.
\begin{equation}
\label{eq:fik-tbl}
    \begin{split}
        C_f = & 
        \frac{4}{R_\delta} \int_0^1 \overline{u}           \dy_{99}+
        4 \int_0^1 \left(y_{99}-1\right)   \overline{u'v'} \dy_{99}- 
        2 \int_0^1 \left(y_{99}-1\right)^2 I_x  \dy_{99}= \\ &
        \frac{4 (1-\delta_d)}{R_\delta} +
        4 \int_0^1 \left(y_{99}-1\right)   \overline{u'v'} \dy_{99}- 
        2 \int_0^1 \left(y_{99}-1\right)^2 I_x  \dy_{99},
    \end{split}
\end{equation}
where $R_\delta=\delta_{99}^* U_\infty^*/\nu^*$ and the definition of displacement thickness, $\delta_d = \int_0^1 (1-\overline{u}) \dy_{99}$, has been used.

\begin{figure}
    \begin{subfigure}{0.5\linewidth}
    \includegraphics[width=\linewidth]{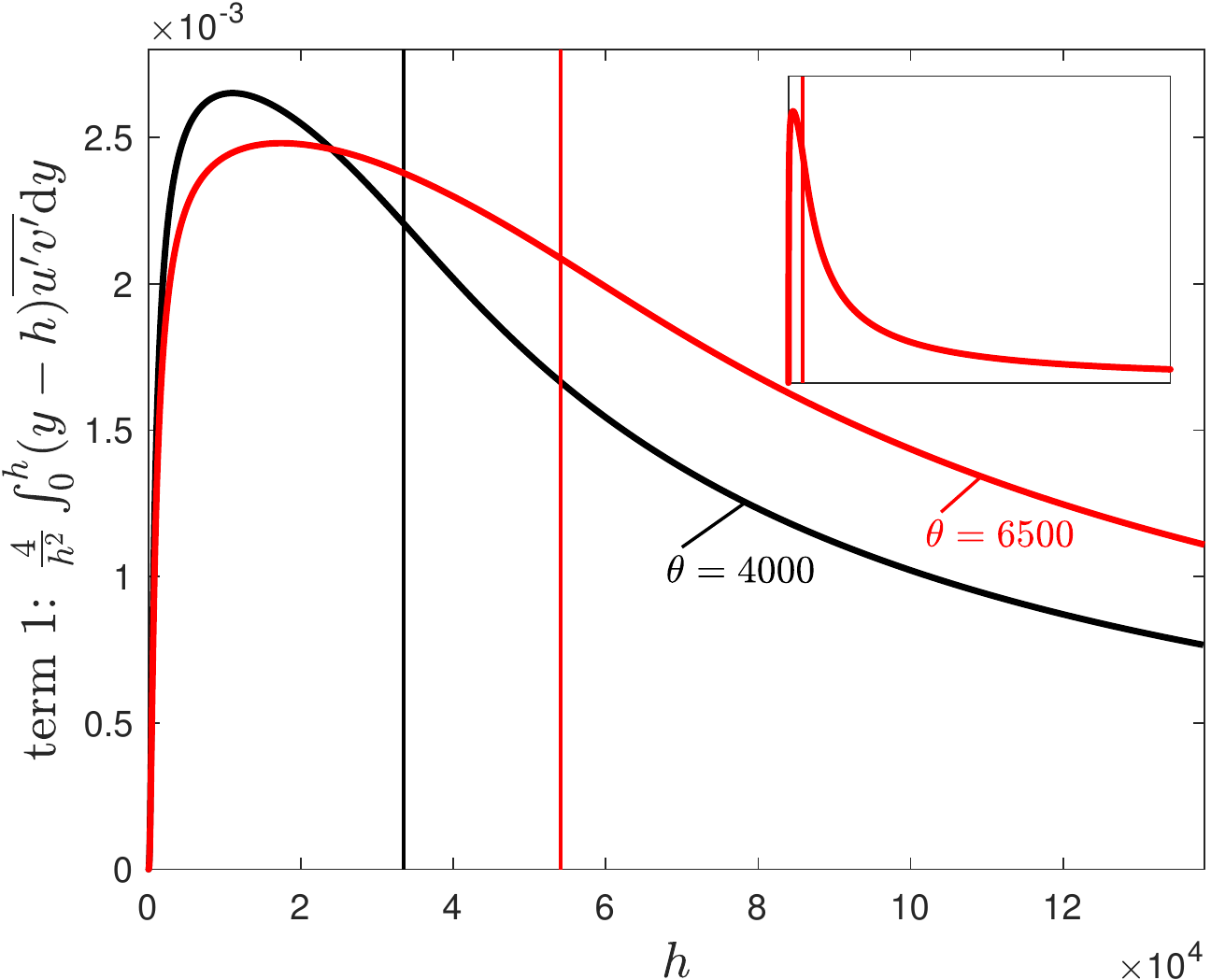}
    \caption{}
    \end{subfigure}
    \begin{subfigure}{0.5\linewidth}
    \includegraphics[width=\linewidth]{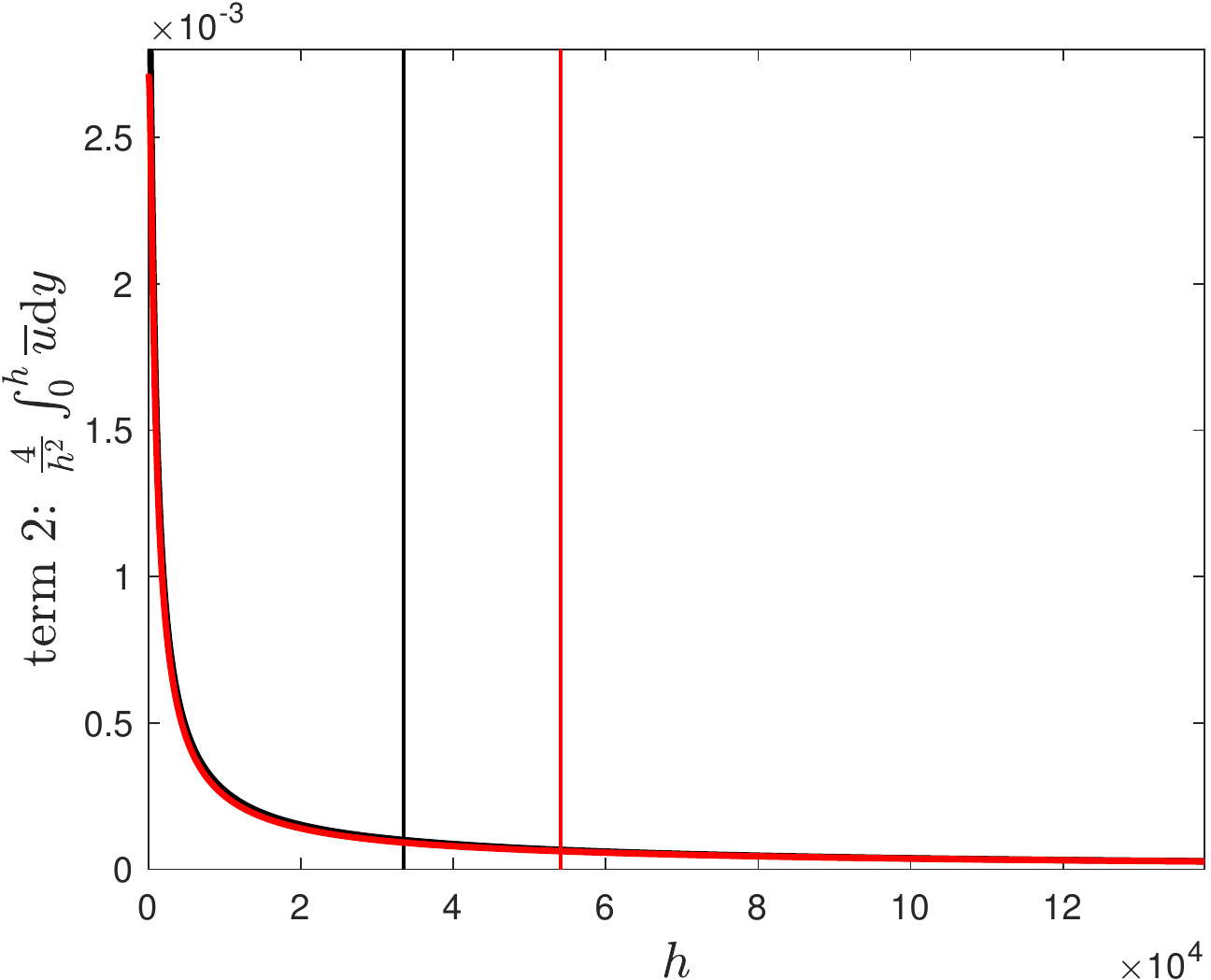}
    \caption{}
    \end{subfigure}
    \caption{Dependence of term 1 (graph a) and term 2 (graph b) in \eqref{eq:int-1-2} on the upper integration bound $h$ for free-stream boundary layers at two Reynolds numbers. The inset of graph (a) shows the decay of term 1 at large $h$ values. In this figure and in figure \ref{fig:terms-3-4-5}, the data are from the direct numerical simulations of \cite{sillero2013one} and the vertical lines indicate the wall-normal locations where $h=\delta_{99}$.}
    \label{fig:terms-1-2} 
\end{figure}

\begin{figure}
    \begin{subfigure}{0.5\linewidth}
    \includegraphics[width=\linewidth]{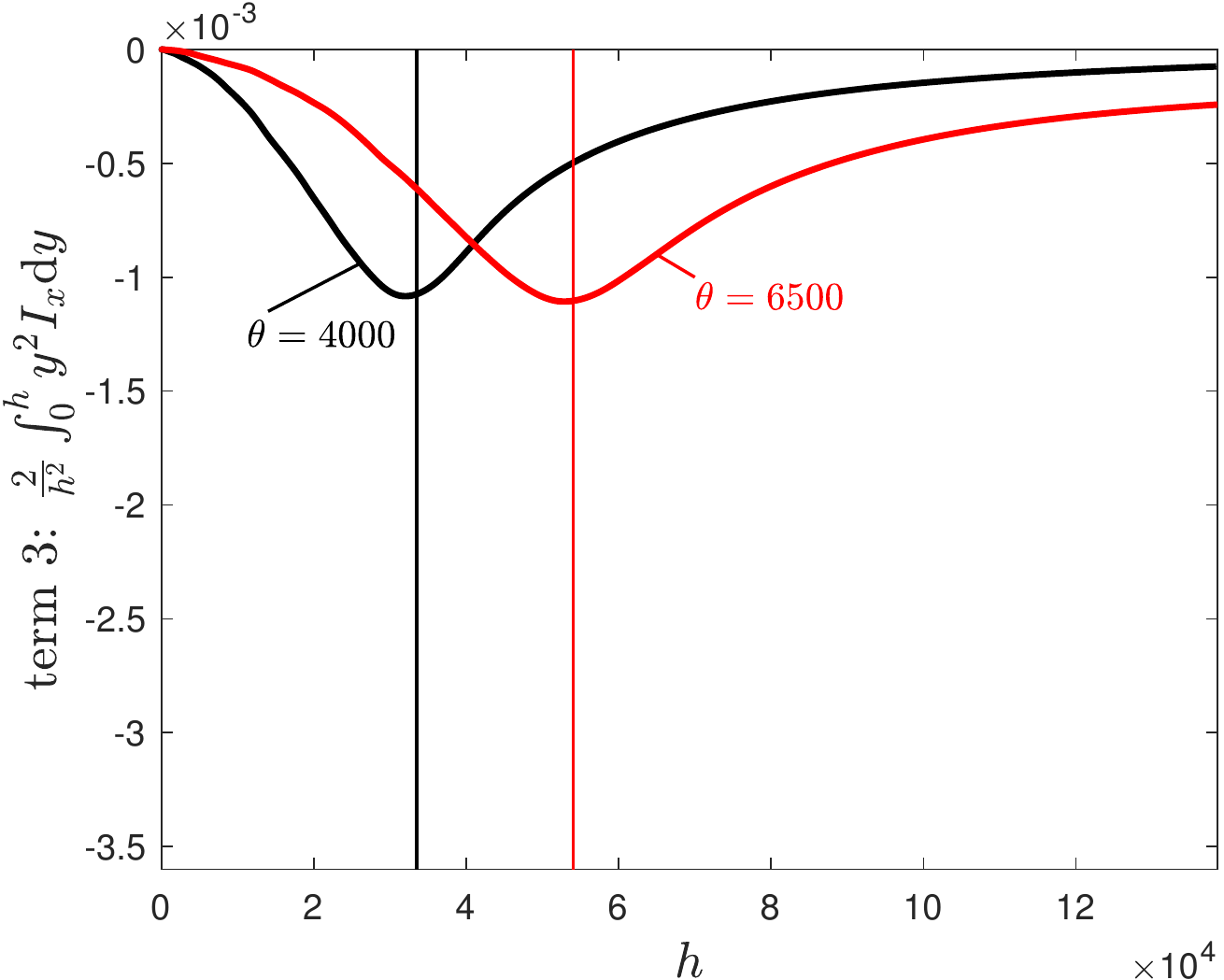}
    \caption{}
    \end{subfigure}
    \begin{subfigure}{0.5\linewidth}
    \includegraphics[width=\linewidth]{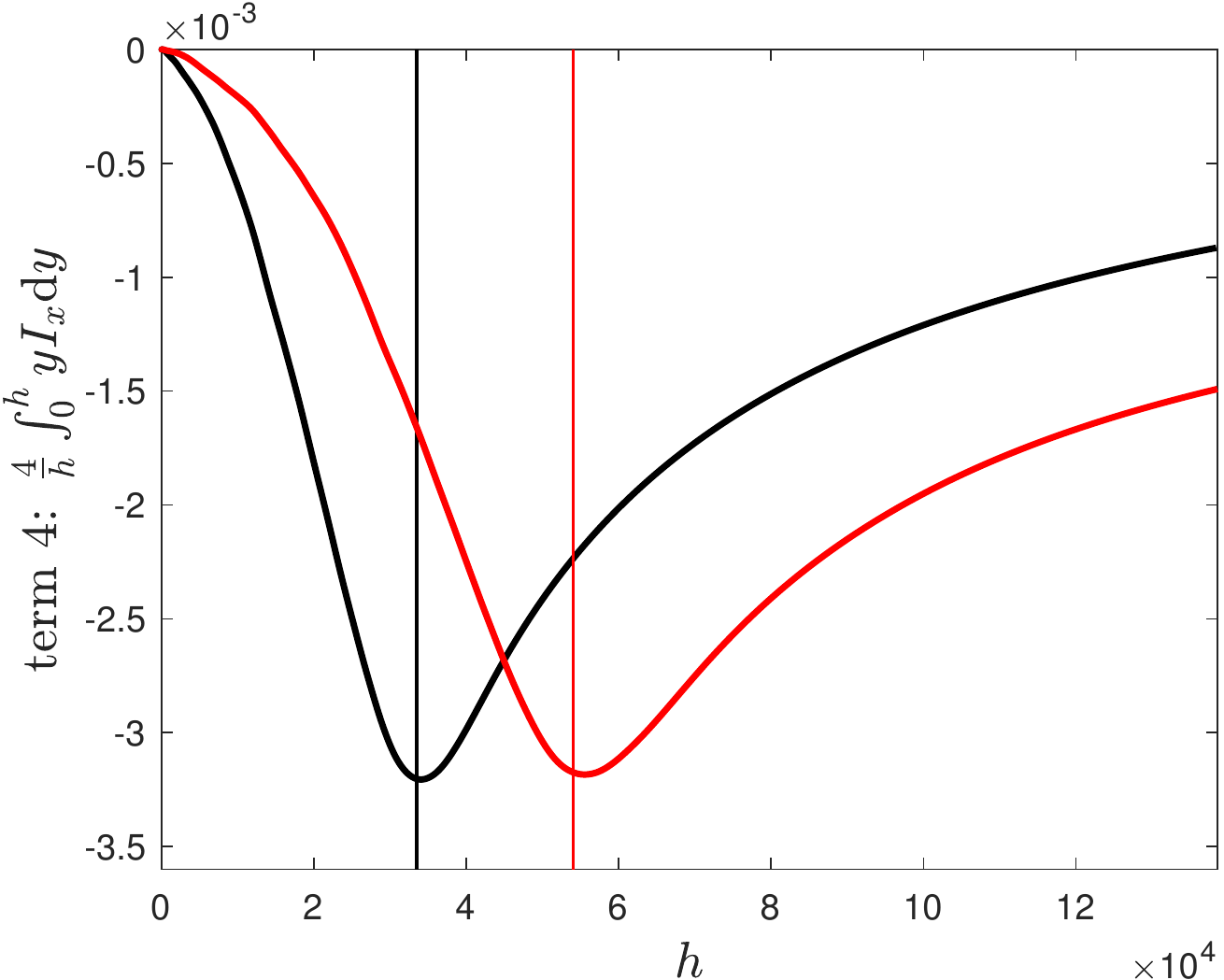}
    \caption{}
    \end{subfigure}    
    \begin{subfigure}{0.5\linewidth}    
    \includegraphics[width=\linewidth]{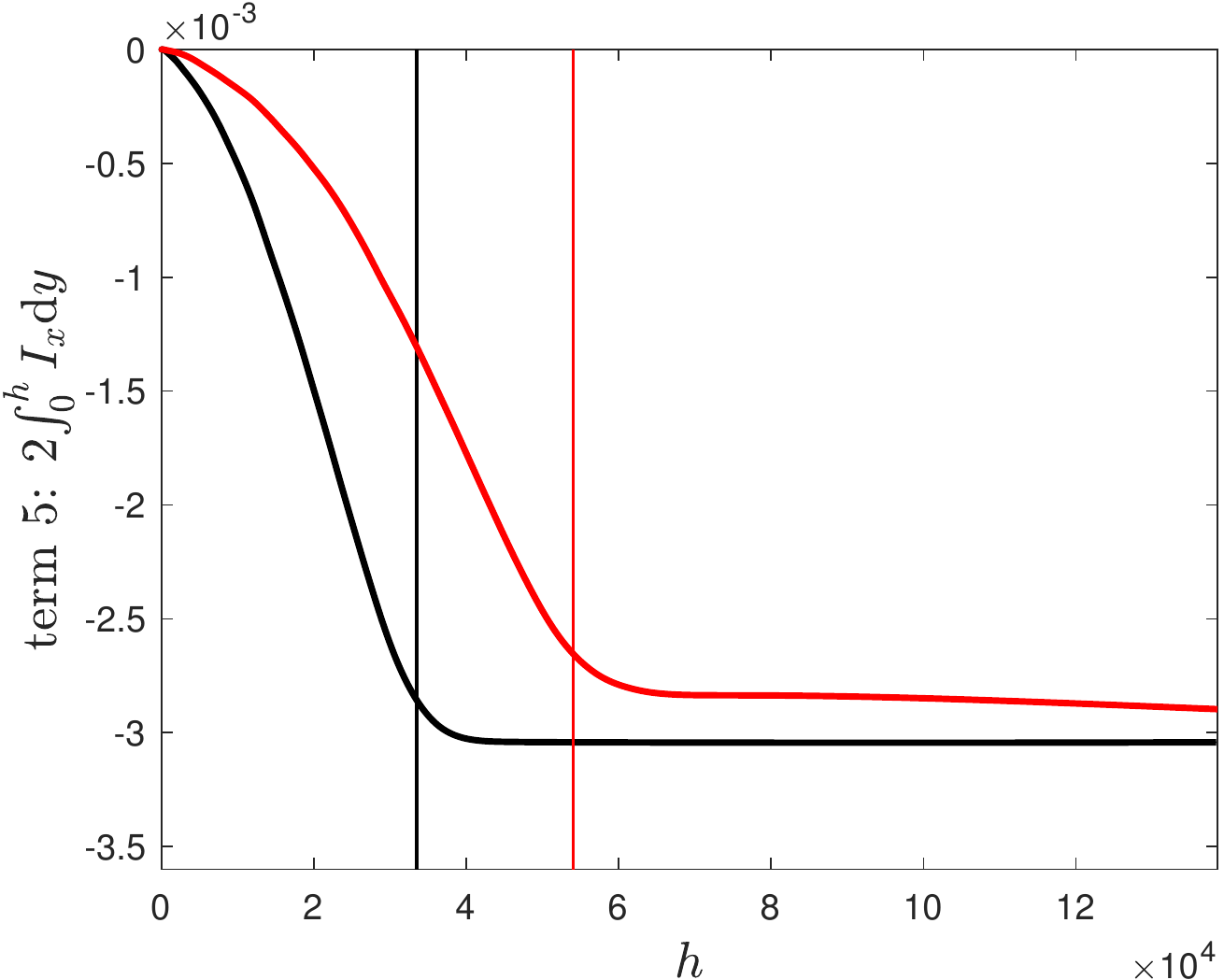}
    \caption{}
    \end{subfigure}
    \caption{Dependence of terms 3, 4, 5 in \eqref{eq:int-3-4} (a,b,c, respectively) on the upper integration bound $h$ for free-stream boundary layers at two Reynolds numbers.}
    \label{fig:terms-3-4-5} 
\end{figure}

The terms on the right-hand side of \eqref{eq:int-1-2} must not depend on the integration bound $h$ because the skin-friction coefficient on the left-hand side does not. The only requirement is that the integration be conducted up to a sufficiently large location for the mean-flow velocity to match the free-stream flow $\{U_\infty^*,0,0\}$. The bound $h$ can therefore be taken asymptotically large. By comparing the integration bounds in the original FIK identity \eqref{eq:fik-tbl} with those in \eqref{eq:int-1-2}, it is evident that the choice of scaling $y^*$ with $\nu^*/U_\infty^*$ instead of $\delta_{99}^*$ allows us to perform the limit $h$$\rightarrow$$\infty$.
In the limit $h$$\rightarrow$$\infty$, term 1 in \eqref{eq:int-1-2} is null as the integral involving the Reynolds stresses is finite because $\overline{u'v'}$ is null in the free stream and term 2 in \eqref{eq:int-1-2} is null because the integral grows $\sim$$h$ as $y$$\rightarrow$$\infty$ because $\overline{u}$$\rightarrow$$1$. Figure \ref{fig:terms-1-2} shows the dependence of terms 1 and 2 on $h$. Term 1 in figure \ref{fig:terms-1-2}(a) decays to zero for an $h$ value that is much larger than the boundary-layer thickness because of the growth of $y-h$ inside the integral, although $\overline{u'v'}$ is mostly contained within the boundary layer. 
It follows that
\begin{equation}
\label{eq:int-3-4}
C_f = 
- \lim_{h \rightarrow \infty} 
\left[
\underbrace{\frac{2}{h^2} \int_0^h y^2 I_x \dy}_{\mbox{term 3}}
- \underbrace{\frac{4}{h} \int_0^h y I_x \dy}_{\mbox{term 4}}
+ \underbrace{2 \int_0^h I_x \dy}_{\mbox{term 5}}
\right].
\end{equation}
Only term 5 in \eqref{eq:int-3-4} is finite as $h$$\rightarrow$$\infty$ because terms 3 and 4 in \eqref{eq:int-3-4} are null in this limit as their integrals are finite because $I_x$ is null in the free stream. The graphs (a), (b), (c) of figure \ref{fig:terms-3-4-5} display the change of terms 3, 4 and 5 with $h$. Terms 3 and 4 show an intense dependence on $h$ for $h$ values comparable to the boundary-layer thickness, although term 5 plateaus to a constant value as soon as the integration is performed up to the free stream.

\begin{figure}
    \centering
    \includegraphics[width=0.7\linewidth]{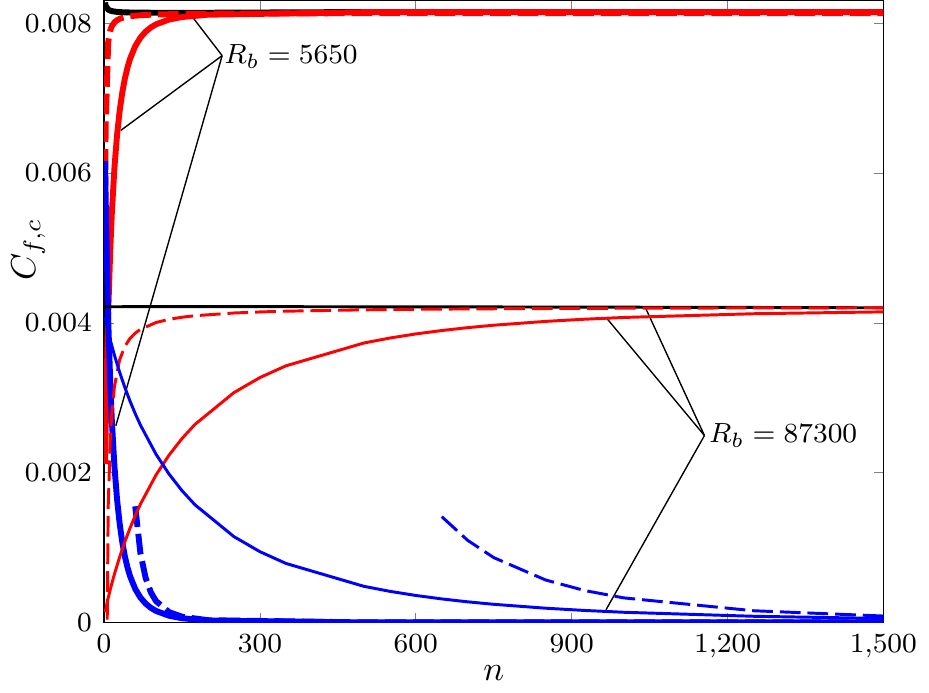}
    \caption{Dependence of the skin-friction terms of \eqref{eq:fik-channel-alternative} (solid lines) and the asymptotic results of \eqref{eq:watson} (dashed lines) on the iterations $n$ for $R_b$$=$$5650$ (thick lines, $R_\tau$$=$$u_\tau^* h_c^*/\nu^*$$=$$180$, where $u_\tau^*$ is the wall-friction velocity) and for $R_b$$=$$87300$ (thin lines, $R_\tau$$=$$2004$). The black, blue and red lines indicate $C_f$, the terms depending on $\overline{u'v'}$ and the terms depending on $\overline{u}$, respectively. The solid lines are computed using the direct numerical simulation data of \cite{hoyas-jimenez-2006}.}
    \label{fig:channel} 
\end{figure}

Equation \eqref{eq:int-3-4} therefore simplifies to \eqref{eq:fik}, which proves that, in the case of a free-stream boundary layer, the FIK identity reduces to the von K\'arm\'an momentum equation between the skin-friction coefficient and the momentum thickness, equation \eqref{eq:theta-2}. 
The identity therefore loses its power of revealing the contribution of the different terms of the $x$-momentum equation to the wall friction. Most notably, the Reynolds stresses disappear from the identity. In the derivation of the FIK identity in channel or pipe flows, no ambiguity exists about the integration bounds, which are fixed by the walls and the centreline in the channel-flow case or the pipe axis in the pipe-flow case. In the case of a free-stream boundary layer, the upper bound of integration is instead not defined by the system geometry because the flow is unconfined. If a finite $h$ value is used as the upper integration bound, as performed in \cite{fukagata-iwamoto-kasagi-2002} and subsequent studies where the boundary-layer thickness $\delta_{99}^*$ was chosen, the contributions of the different terms to the wall friction depend on $h$. However, this dependence is spurious because their influence on the skin-friction coefficient must obviously be independent of $h$. 
When $h=\delta_{99}$, one may be led to confirm the established result that the Reynolds stresses impact significantly on the wall-shear stress by noting that the Reynolds-stress term 1 is comparable to the skin-friction coefficient ($C_f=3.03 \times 10^{-3}$ for $\theta=4000$ and $C_f=2.71 \times 10^{-3}$ for $\theta=6500$), as shown in figure \ref{fig:terms-1-2}(a). However, the non-physical dependence of term 1 on $h$ precludes the quantification of the effect of the Reynolds stresses on the wall friction.

\cite{xia-etal-2015} and \cite{wenzel-etal-2022} performed only two wall-normal integrations, instead of three as in \cite{fukagata-iwamoto-kasagi-2002}, stating that a twofold repeated integration is more suitable for a physical interpretation. \cite{wenzel-etal-2022}'s equation (3.7) in the zero-Mach-number limit coincides with our \eqref{eq:fik-2} by setting $y$$=$$h$. Similarly to \eqref{eq:int-1-2}, the twofold-integration identity also shows the spurious dependence on $h$ and reduces to the von K\'arm\'an momentum equation \eqref{eq:theta-2} as $h$$\rightarrow$$\infty$. 

\cite{sbragaglia-sugiyama-2007} proved that, in the case of channel and pipe flows, the weighing function $1-y$ in the integral involving the Reynolds stresses in the FIK identity can be interpreted physically as the velocity gradient of the corresponding Stokes-flow solution (this result was also used by \cite{modesti-etal-2018}). As the corresponding Stokes-flow solution cannot be obtained in the case of free-stream boundary layers, \cite{sbragaglia-sugiyama-2007}'s result confirms our finding that the Reynolds-stress integral in \eqref{eq:int-1-2} does not possess a precise physical meaning for free-stream boundary layers.

\subsection{Alternative FIK identities}
\label{sec:fik-2}

\cite{bannier-etal-2015} remarked that a third integration along $y$ could be performed before the final integration \eqref{eq:fik-3} up to $y=h$, thereby obtaining an alternative FIK identity. As shown by \cite{wenzel-etal-2022}, an infinite number $n$ of successive integrations between 0 and $y$ can in fact be performed before the final integration between 0 and $h$. The result is
\begin{equation}
\label{eq:fik-alternative-1}
    \begin{split}
        C_f = 
        -\frac{2n}{h^n}\int_0^h (h-y)^{n-1} \overline{u'v'} \dy + 
        \frac{2n(n-1)}{h^n}\int_0^h (h-y)^{n-2} \overline{u} \dy -
        \frac{2}{h^n} \int_0^h (h-y)^n I_x \dy.
        \end{split}
\end{equation}
The identities \eqref{eq:fik-alternative-1} are valid for $n\geq2$. For $n=2$, \eqref{eq:fik-alternative-1} is \eqref{eq:int-1-2}. For every $n$, the identities \eqref{eq:fik-alternative-1} simplify to \eqref{eq:fik} as $h$$\rightarrow$$\infty$. In this limit, the first term on the right-hand side of \eqref{eq:fik-alternative-1} is null because $h^n$ appears at the denominator and the integral is finite, and the second term is null because the integral always grows more slowly than the denominator $h^n$. The third term in \eqref{eq:fik-alternative-1} is expanded by using the binomial theorem,
\begin{equation}
\label{eq:binomial}
    \begin{split}
        -\frac{2}{h^n} \int_0^h (h-y)^n I_x \dy=
        - 2 \sum_{k=0}^n \binom{n}{k} \frac{(-1)^k}{h^k} \int_0^h y^k I_x \dy.
    \end{split}
\end{equation}
As $h$$\rightarrow$$\infty$, the terms on the right-hand side of \eqref{eq:binomial} for $k \neq 0$ vanish because the integrals are finite, while the term for $k=0$ in \eqref{eq:binomial} is finite because it is independent of $h$. This remaining term is \eqref{eq:fik}.
Further alternative formulas are found by multiplying \eqref{eq:fik-1} by $y^m$ ($m$$>$$0$) before performing the subsequent integrations and again the final result is \eqref{eq:fik} in the limit $h$$\rightarrow$$\infty$. The existence of alternatives to the original FIK identity for finite $h$ and the simplification of all of them to the von K\'arm\'an momentum equation \eqref{eq:theta-2} further raises questions on the validity of this approach for the quantification of the physical roles of the terms in \eqref{eq:rans-x} on the skin-friction coefficient because the weighed influence of the terms in \eqref{eq:fik-alternative-1} depends on $n$. This dependence on $n$ is spurious because $n$ is not a physical parameter.

Identities analogous to \eqref{eq:fik-alternative-1} can be found for confined flows. For fully developed channel flows, one finds
\begin{equation}
\label{eq:fik-channel-alternative}
    \begin{split}
    \frac{C_{f,c}}{8 (n+1)} = 
    -\int_0^1 (1-y_c)^{n-1} \overline{u_c'v_c'} \dy_c +
    \frac{n-1}{R_b} \int_0^1 (1-y_c)^{n-2} \overline{u}_c \dy_c, 
    \end{split}
\end{equation}
where $y_c=y^*/h_c^*$, $h_c^*$ is the half-channel height, the velocity components are scaled by $2U_b^*$, where $U_b^*$ is the bulk velocity, $R_b=2U_b^*h_c^*/\nu^*$ and $C_{f,c}=(8/R_b) \mathrm{d} \overline{u}_c/{\mathrm{d}y_c}|_{y_c=0}$.
The identity \eqref{eq:fik-channel-alternative} is valid for $n \geq 2$. The identity found by \cite{fukagata-iwamoto-kasagi-2002} is obtained for $n$$=$$2$ (they integrate to the upper wall in their (16)). In the laminar case, for which $\overline{u_c'v_c'}=0$ and $\overline{u}_c=3y_c(2-y_c)/4$, \eqref{eq:fik-channel-alternative} is independent of $n$ as the term containing $\overline{u}_c$ simplifies and the identity reduces to the laminar $C_{f,c}=12/R_b$. Amongst the $n$-family of identities \eqref{eq:fik-channel-alternative}, only the identity obtained by \cite{fukagata-iwamoto-kasagi-2002}, found for $n$$=$$2$, possesses a clear physical meaning in the turbulent-flow case because the term involving the mean velocity $\overline{u}_c$ in \eqref{eq:fik-channel-alternative} reduces to the part of the skin-friction coefficient that pertains to a laminar channel flow by using the definition of bulk velocity (this distinction does not emerge directly in the case of a turbulent boundary layer as the wall friction of the Blasius boundary layer is not retrieved in a single term in \eqref{eq:int-1-2}, as pointed out by \cite{fukagata-iwamoto-kasagi-2002}). For $n$$=$$2$, the term involving $\overline{u_c'v_c'}$ in \eqref{eq:fik-channel-alternative} therefore univocally distils the effect of the turbulence on the skin-friction coefficient. For $n$$>$$2$, the term containing $\overline{u}_c$ cannot be simplified and the laminar and turbulent contributions to the skin-friction coefficient cannot be distinguished.

In order to study the asymptotic behaviour of \eqref{eq:fik-channel-alternative} as $n$$\rightarrow$$\infty$, we write \eqref{eq:fik-channel-alternative} as
\begin{equation}
\label{eq:fik-channel-alternative-2}
    C_{f,c} = 
    - 8 (n+1) \int_0^\infty \overline{u_c'v_c'}    e^{-ns}        \mathrm{d}s +
    \frac{8 (n^2-1)}{R_b} \int_0^\infty \overline{u}_c e^s e^{-ns}    \mathrm{d}s,
\end{equation}
where $s$$=$$-$$\ln(1-y_c)$. Appendix \ref{app:b} shows that the limit of the integrals in \eqref{eq:fik-channel-alternative-2} as $n$$\rightarrow$$\infty$ can be moved inside the integrals because the integrands converge uniformly. We expand $\overline{u_c'v_c'} \sim s^{\alpha_1} \sum_{k=0}^\infty a_{1k} s^{k \beta_k}$ as $s \rightarrow 0^+$,
\begin{equation}
\begin{split}
    \overline{u_c'v_c'} \sim & 
    A_{uv3} y_c^3 + A_{uv4} y_c^4 + \mathcal{O}\left(y_c^5\right)= 
    A_{uv3} \left(1-e^{-s}\right)^3 + A_{uv4} \left(1-e^{-s}\right)^4 + ...= \\ &
    s^3 \left[A_{uv3} + \left(A_{uv4} - \frac{3A_{uv3}}{2}\right)s\right]
    +\mathcal{O}\left(s^5\right),
\end{split}
\end{equation}
where $A_{uv3}(R_b)$ and $A_{uv4}(R_b)$ are determined numerically. We expand $\overline{u}_c e^s \sim s^{\alpha_2} \sum_{k=0}^\infty a_{2k} s^{k \beta_2}$ as $s$$\rightarrow$$0^+$, 
\begin{equation}
    \begin{split}
    \overline{u}_c e^s \sim &
    \left[ A_{\overline{u}1} y_c + A_{\overline{u}2} y_c^2 + A_{\overline{u}3} y_c^3 + A_{\overline{u}4} y_c^4
    + \mathcal{O}\left(y_c^5\right) \right] e^s = \\ &
    A_{\overline{u}1} \left(e^s-1\right) +
    A_{\overline{u}2} \left(e^s + e^{-s}-2\right) + 
    A_{\overline{u}3} \left(e^s + 3 e^{-s} - e^{-2s} - 3\right) + \\ &
    A_{\overline{u}4} \left(e^s + 6 e^{-s} - 4 e^{-2s} + e^{-3s} - 4\right) +
    ...= \\ &
    s\left[
        A_{\overline{u}1} + \left( \frac{A_{\overline{u}1}}{2} + A_{\overline{u}2}\right)s +
        \left( \frac{A_{\overline{u}1}}{3} + A_{\overline{u}3}\right)s^2 +
        \left( \frac{A_{\overline{u}1}}{24} + \frac{A_{\overline{u}2}}{12} 
        - \frac{A_{\overline{u}3}}{2} + A_{\overline{u}4} \right)s^3
    \right]
    +\mathcal{O}\left(s^5\right),
    \end{split}
    \end{equation}
where $A_{\overline{u}1}={\mathrm{d}\overline{u}/\dy_c}|_{y_c=0}$, $A_{\overline{u}2}=0.5{\mathrm{d}^2\overline{u}/\dy_c^2}|_{y_c=0}$, $A_{\overline{u}3}=(1/6){\mathrm{d}^3\overline{u}/\dy_c^3}|_{y_c=0}$ and $A_{\overline{u}4}=(1/24){\mathrm{d}^4\overline{u}/\dy_c^4}|_{y_c=0}$. It follows that $\alpha_1=3$, $\beta_1=1$, $a_{10}=A_{uv3}$, $a_{11}=A_{uv4}-3 A_{uv3}/2$, $\alpha_2=1$, $\beta_2=1$, $a_{20}=A_{\overline{u}1}$, $a_{21}=A_{\overline{u}2}+A_{\overline{u}1}/2$, $a_{22}=A_{\overline{u}3}+A_{\overline{u}1}/3$ and $a_{23}=A_{\overline{u}1}/24 + A_{\overline{u}2}/12 - A_{\overline{u}3}/2 + A_{\overline{u}4}$. 
According to Watson's lemma \citep{bender-orszag-1999}, as $n$$\rightarrow$$\infty$,
\begin{equation}
\label{eq:watson}
\begin{split}
        C_{f,c} 
        \sim 
        &       
        - 8 (n+1) \left[ 
                \frac{\Gamma(4)A_{uv3}}{n^4} 
                +
                \left(A_{uv4} - \frac{3 A_{uv3}}{2}  \right) \frac{\Gamma(5)}{n^5}
                + ...
            \right] + \\ &
            \frac{8 (n^2-1)}{R_b} 
            \left[ 
                \frac{\Gamma(2)A_{\overline{u}1}}{n^2} 
                + \left( A_{\overline{u}2} + \frac{A_{\overline{u}1}}{2} \right)
                \frac{\Gamma(3)}{n^3}
                + 
                \left( A_{\overline{u}3} + \frac{A_{\overline{u}1}}{3} \right)
                \frac{\Gamma(4)}{n^4} \right.
                + \\ &
                \left.
                \left( A_{\overline{u}4} - \frac{A_{\overline{u}1}}{2} + \frac{A_{\overline{u}2}}{12} + \frac{A_{\overline{u}1}}{24} \right)
                \frac{\Gamma(5)}{n^5}
                +
                ...
            \right]
            \sim
            \frac{8}{R_b} \frac{\mathrm{d}\overline{u}}{\dy_c}\biggm|_{y_c=0},         
    \end{split}
\end{equation}
where $\Gamma$ is the gamma function. The asymptotic analysis is useful because it proves that, as $n$ grows, the integral in \eqref{eq:fik-channel-alternative} involving the Reynolds stresses impacts less and less on the skin-friction coefficient because it behaves $\sim - 48 A_{uv3}/n^3$, while the term containing the mean flow becomes more and more relevant because it behaves $\sim (8/R_b) \mathrm{d}\overline{u}/\dy_c|_{y_c=0} + 4\left( {\mathrm{d}^2\overline{u}/\dy_c^2}|_{y_c=0} + {\mathrm{d}\overline{u}/\dy_c}|_{y_c=0} \right)/(R_b n)$. Figure \ref{fig:channel} shows the skin-friction terms as functions of $n$ at two Reynolds numbers, computed numerically via \eqref{eq:fik-channel-alternative} and asymptotically via \eqref{eq:watson}. As $n$$\rightarrow$$\infty$, no information on the physics of a turbulent channel flow emerges from \eqref{eq:fik-channel-alternative} as the Reynolds stresses vanish and the identity degenerates to the definition of the skin-friction coefficient, $C_{f,c}=(8/R_b) \mathrm{d} \overline{u}_c/{\mathrm{d}y_c}|_{y_c=0}$. The asymptotic behaviour \eqref{eq:watson} further proves that the channel-flow identity \eqref{eq:fik-channel-alternative} only possesses a defined physical meaning when $n=2$.

Motivated by the studies of \cite{xia-etal-2015} and \cite{wenzel-etal-2022} on free-stream boundary layers, we perform a twofold integration in the fully developed channel-flow case. The result is
\begin{equation}
\label{eq:fik-channel-two-fold}
    C_{f,c} = \frac{16}{R_b} \overline{u}_c(y_c=1) - 16 \int_0^1 \overline{u_c'v_c'} \dy_c. 
\end{equation}
The identity \eqref{eq:fik-channel-two-fold} reduces to $C_{f,c}=12/R_b$ in the laminar case, i.e. when $\overline{u_c'v_c'}=0$ and $\overline{u}_c$$=$$3/4$ at the centreline. Differently from the original FIK identity, relation \eqref{eq:fik-channel-two-fold} lacks the virtue of univocally distinguishing the laminar and the turbulent contributions to the skin-friction coefficient because $\overline{u}_c(y_c=1)$ is the mean velocity at the centreline. Nevertheless, it can be useful for checking numerical calculations and experimental measurements of $C_f$, computed directly via the wall-normal velocity gradient at the wall or the mean streamwise pressure gradient, and indirectly via the $\overline{u_c'v_c'}$ profile and the mean centreline velocity. 
It is found that \eqref{eq:fik-channel-two-fold} is also valid for pipe flows, in which case $y_c=r^*/R^*$, $r^*$ is the radial coordinate, $R^*$ is the pipe radius and $R_b=2 U_b^* R^*/\nu^*$ ($C_{f,c}=16/R_b$ is found in the laminar case as $\overline{u_c'v_c'}=0$ and $\overline{u}_c$$=$$1$ at the pipe axis).
As the Reynolds number increases, it is progressively more difficult to measure the wall-shear stress via direct measurement of the wall-normal velocity gradient at the wall because the near-wall turbulent length scales become smaller and the viscous sublayer thinner. In the limit of large Reynolds number, it is instead easier to compute the skin-friction coefficient via \eqref{eq:fik-channel-two-fold} because the measurements of the bulk velocity and the integrated Reynolds stresses suffer progressively less from the large near-wall velocity gradients. Furthermore, the identity \eqref{eq:fik-channel-two-fold} allows for a local skin-friction measurement, while computing the wall-shear stress via the streamwise pressure gradient may require wall-pressure measurements distributed along a long streamwise stretch. These comments are also valid for the original identities by \cite{fukagata-iwamoto-kasagi-2002}. During the final revision stages of the present work, we became aware that \eqref{eq:fik-channel-two-fold} was also discovered by \cite{elnahhas-johnson-2022} for channel flows.

The FIK identity for planar Couette flow was obtained by \cite{kawata-alfredsson-2019}. It is worth noting that, in that case, the laminar and turbulent contributions to the skin-friction coefficients were distinguished by integrating twice, while such a result is attained by integrating thrice in the case of channel and pipe flows.

\subsection{Skin-friction coefficient as a function of integral thicknesses}
\label{sec:rd}

After verifying that the FIK identity \eqref{eq:int-1-2} simplifies to the von K\'arm\'an momentum equation \eqref{eq:theta-2}, we follow the study of \cite{renard2016theoretical}, who obtained an integral identity for free-stream boundary layers where the interval of integration is unbounded. The central idea is to decompose the momentum thickness \eqref{eq:momentum-thickness} as the sum of integral thicknesses in order to quantify the impact of each term in the mechanical energy balance on the skin-friction coefficient, via \eqref{eq:theta-2}, and on the wall-friction drag, via \eqref{eq:drag}.
Differently from \cite{renard2016theoretical}, we do not adopt the boundary-layer approximation, i.e. the term ${\p^2 \overline{u}}/{\p x^2}$ is kept in the $x$-momentum equation \eqref{eq:rans-x}. We multiply \eqref{eq:rans-x} by $\ou-1$ and integrate along $y$ from 0 to $\infty$ to find

\begin{equation}
\label{eq:cf-int}
{C_f}=2\int_0^{\infty} \frac {\p \ou}{\p y} \left(\frac {\p \ou}{\p y}+\frac{\p \ov}{\p x} \right)\dy+2\int_0^{\infty} -\overline{u'v'}\frac{\p \ou}{\p y}\dy+2\int_0^{\infty} -\ou~\ov\frac {\p \ou}{\p y}\dy + 2\int_0^{\infty} (\ou-1)\frac{\p \ou^2}{\p x}\dy,
\end{equation}
which may be written as
\begin{equation}
\label{eq:cf-parts}
C_f=\mathcal{E}+\mathcal{P}+\C+\SSc.
\end{equation}

\noindent
The five terms in \eqref{eq:cf-parts} can be interpreted from the perspective of an energy balance (per unit time), by multiplying \eqref{eq:cf-parts} by $\rho^* U_\infty^{*3}$, or, as a force balance, by multiplying \eqref{eq:cf-parts} by $\rho^* U_\infty^{*2}$. In the former case, the meaning of the terms is clear if the absolute frame of reference is adopted, i.e. where the wall moves and the free stream is stationary \citep{renard2016theoretical}. The left-hand side is the energy imparted by the moving wall on the fluid, while the first term on the right-hand side is the energy dissipated into heat by the viscous action of the mean flow, and the second term is the energy spent on creating turbulence. The third and fourth terms represent the uptake of kinetic energy of the fluid by the moving wall and are related to the growth of the boundary layer. The convection term ($\C$) is negative, which explains why blowing through the wall (positive $\ov$) decreases the drag, while suction (negative $\ov$) increases the drag. The fourth term ($\SSc$) is named the streamwise-heterogeneity term \citep{fan-etal-2020}. 

\noindent
In order to interpret the terms
\begin{equation}
\label{eq:pepsilon}
\mathcal{E}+\mathcal{P}=2\int_0^{\infty} \frac {\p \ou}{\p y} \left(\frac {\p \ou}{\p y}+\frac{\p \ov}{\p x} \right)\dy+2\int_0^{\infty} -\overline{u'v'}\frac{\p \ou}{\p y}\dy,
\end{equation}
we multiply \eqref{eq:rans-x} by $\ou$ and integrate from zero to $\infty$. Following \cite{schlichting-gersten-2003}, we obtain

\begin{equation}
\label{eq:energy}
\mathcal{E}+\mathcal{P}=\frac{\dd \Eb}{\dx}
\end{equation}
where
\begin{equation}
\label{eq:E}
\Eb=\int_0^{\infty} \ou \left(1-\ou^2\right)\dy
\end{equation}
is the energy thickness. 
Note that $\mathcal{E}$ may be written as $2/\Delta$, where $\Delta$ is the dissipation thickness \citep{hinze-1975}, which is
\begin{equation}
\label{eq:dissth}
\Delta=\left[\int_0^{\infty} \left(\frac {\p \ou}{\p y}\right)^2\dy \right]^{-1}
\end{equation}
when the boundary-layer approximation ($\p \ov/\p x \ll \p \ou/\p y $) is used in \eqref{eq:pepsilon}.

\noindent
For the convection term
\begin{equation}
\label{eq:C1}
\C=2\int_0^{\infty} -\ou \ \ov\frac {\p \ou}{\p y}\dy,
\end{equation}
we use continuity and integration by parts to find
\begin{equation}
\label{eq:C2}
\C= 2\int_0^{\infty} \ou\frac {\p \ou}{\p y}\int_0^{\hat{y}} \frac{\p \ou}{\p x} \mathrm{d}\hat{y} \dy = \int_0^{\infty} \left(1-\ou^2\right) \frac{\p \ou}{\p x} \dy.
\end{equation}
Equation \eqref{eq:C2} can be written as
\begin{equation}
\label{eq:Cfinal}
\C=\frac{\dd \Cb}{\dx} \quad \mbox{where} \quad \Cb=\int_0^{\infty} \ou \left (1-\frac{1}{3}\ou^2\right)\dy.
\end{equation}

\begin{figure}
    \begin{subfigure}{0.5\linewidth}
    \includegraphics[width=\linewidth, clip=true]{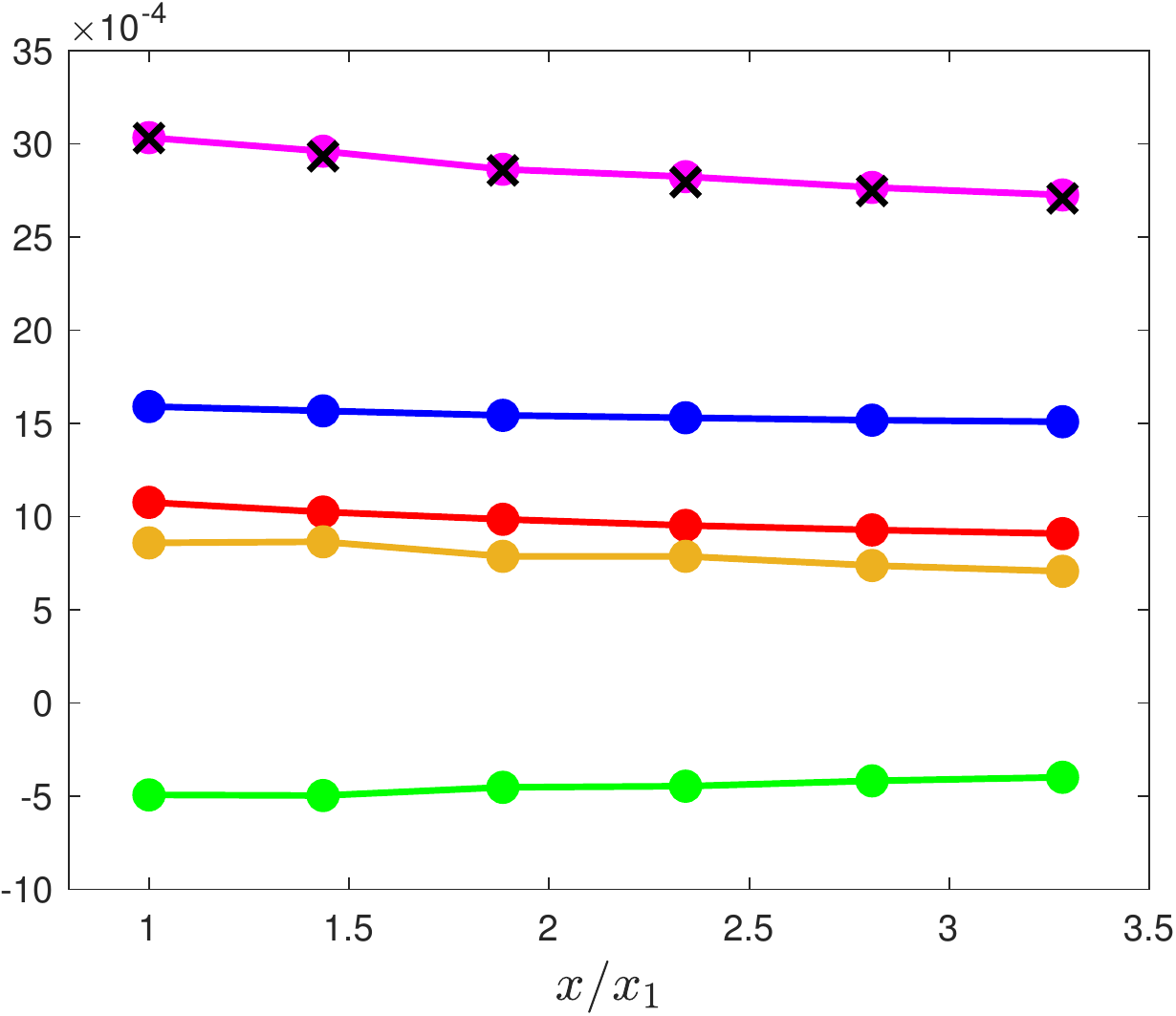}
    \label{fig:RD-sillero} 
    \vspace{-0.3cm}
    \caption{}
    \end{subfigure}
    \begin{subfigure}{0.5\linewidth}
    \includegraphics[width=\linewidth, clip=true]{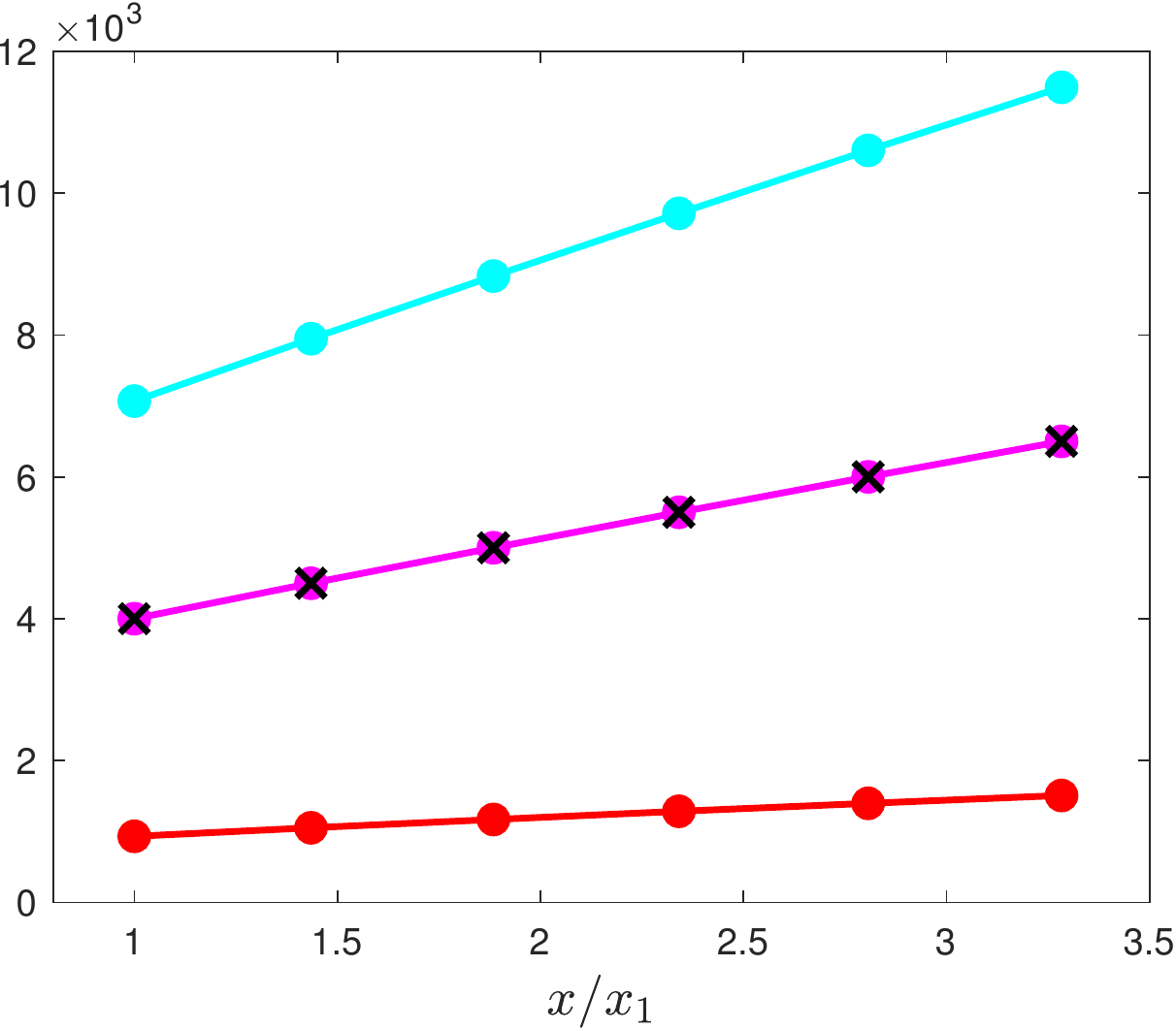}
    \label{fig:thickness-sillero} 
    \vspace{-0.3cm}
    \caption{}
    \end{subfigure}
    \caption{(a) Decomposition of the skin-friction coefficient $C_f$ into the terms in \eqref{eq:cf-parts}: $\mathcal{E}$ (red \textcolor{red}{{$\medbullet $}}), $\mathcal{P}$ (blue \textcolor{blue}{{$\medbullet $}}), $\C$ (green \textcolor{green}{{$\medbullet $}}) and $\SSc$ (orange \textcolor{orange}{{$\medbullet $}}). The magenta symbol \textcolor{magenta}{{$\medbullet $}} indicates the sum of all four components on the right-hand side of \eqref{eq:cf-parts}. The black crosses indicate $C_f$ obtained directly from the wall-shear stress data of \cite{sillero2013one}. (b) Integral lengths in \eqref{eq:theta}: energy thickness $\Eb$ (cyan \textcolor{cyan}{{$\medbullet $}}), inertia thickness $\Ib$ (red \textcolor{red}{{$\medbullet $}}) and momentum thickness $\theta$ (black crosses). The magenta symbol \textcolor{magenta}{{$\medbullet $}} indicates $(\Eb+\Ib)/2$. The $x-$axis is scaled by $x_1$, the coordinate of the first point. The momentum thickness $\theta$ at the six points is $4000, 4500, 5000, 5500, 6000, 6500$. The data in this figure are obtained by post-processing the results of the direct numerical simulations of \cite{sillero2013one}.}
    \label{fig:RD} 
\end{figure}

\noindent
For the streamwise-heterogeneity term,
\begin{equation}
\label{eq:S1}
\SSc=2 \int_0^{\infty} (\ou-1)\frac{\p \ou^2}{\p x} \dy,
\end{equation}
we note that
\begin{equation}
\label{eq:S2}
(\ou-1)\frac{\p \ou^2}{\p x}=\frac{\p }{\p x}\left[ \ou^2\left(\frac{2}{3}\ou-1\right)\right].
\end{equation}
Hence
\begin{equation}
\label{eq:Sfinal}
\SSc=\frac{\dd \Sb}{\dx} \quad \mbox{where} \quad \Sb=2\int_0^{\infty} \ou^2\left(\frac{2}{3}\ou-1\right)\dy.
\end{equation}
By adding $\Cb$ and $\Sb$, one finds
\begin{equation}
\label{eq:c_plus_s}
\Ib=\Cb + \Sb=\int_0^{\infty} \ou \left(1-\ou\right)^2\dy,
\end{equation}
which we term the inertia thickness. To summarize, we have,
\begin{equation}
\label{eq:energy-sum}
\mathcal{E}+\mathcal{P}=\frac{\dd \Eb}{\dx}
\text{ and }
\C+\SSc=\frac{\dd \Ib}{\dx}.
\end{equation}
It is verified that
\begin{equation}
\label{eq:theta}
2\theta=\Eb +  \Ib = \Eb +  \Cb + \Sb
\end{equation}
and
\begin{equation}
\label{eq:momentum-integral-expanded}
C_f=2\frac{\dd \theta}{\dx}=\frac{\dd \Eb}{\dx}+\frac{\dd \Ib}{\dx}=\frac{\dd \Eb}{\dx}+\frac{\dd \Cb}{\dx} + \frac{\dd \Sb}{\dx},
\end{equation}
which is therefore a decomposition of the von K\'arm\'an momentum equation \eqref{eq:theta-2}. The terms of \eqref{eq:cf-parts} and the integral lengths $\Eb$, $\Ib$ and $\theta$, extracted from the numerical data of \cite{sillero2013one}, are shown in figure \ref{fig:RD}. The first part of the relation \eqref{eq:theta}, i.e. the decomposition of $\theta$ into $\Eb$ and $\Ib$, was found by \citet{drela-2009} in the context of aerodynamics of vehicles. In \citet{drela-2009}, the term $\Ib$ was not related to the boundary-layer inertia terms, but it was instead linked to the kinetic-energy excess of the wake behind a vehicle.

Similarly to the study of \cite{renard2016theoretical}, equation \eqref{eq:momentum-integral-expanded} can be interpreted in the absolute frame of reference, i.e. where the wall is in motion. Equation \eqref{eq:momentum-integral-expanded} thus describes how the energy given by the wall motion to the fluid, measured by twice the change of $\theta$ with the streamwise direction, is divided into the change of $\Eb$, representing the losses of mean kinetic energy due to the mean-flow viscous dissipation into heat and to the production of turbulence, and the change of $\Ib$, representing the change in convective transport of the mean kinetic energy due to the mean velocity. The change of $\Ib$ can in turn be expressed as the sum of the changes of the thicknesses $\Cb$ and $\Sb$, which represent the change in transport due to the wall-normal mean velocity and the streamwise mean velocity, respectively.
Referring to \eqref{eq:drag}, we can now utilize the streamwise integral  of \eqref{eq:momentum-integral-expanded} to investigate what percentage of the different terms in the RD decomposition contributes to the total drag by taking differences in the corresponding integral thicknesses. 

It is noted, however, that the RD decomposition \eqref{eq:cf-int} and identities emerging from it, such as \eqref{eq:momentum-integral-expanded}, do not distinguish the laminar and the turbulent contributions to the skin-friction coefficient for any flow, while the FIK identity achieves this task for confined flows and the identity discovered by \cite{elnahhas-johnson-2022} does so for free-stream boundary layers. As demonstrated by \cite{renard2016theoretical}, the difference in the skin-friction coefficient between a laminar and a turbulent boundary layer at the same Reynolds number based on $\Delta$ (for which $\mathcal{E}$ is identical) is dominated by $\mathcal{P}$.

\section{Conclusions}
We have shown that the identity discovered by \cite{fukagata-iwamoto-kasagi-2002}, expressing the skin-friction coefficient of free-stream boundary layers as a function of integrated terms of the Reynolds-averaged streamwise momentum equation, simplifies to the von K\'arm\'an momentum integral equation relating the skin-friction coefficient and the momentum thickness. This simplification arises as the upper integration bound along the wall-normal direction is taken asymptotically large. If the upper bound is finite, the weighted contributions of the terms of the streamwise momentum equation depend spuriously on the bound itself. The family of infinite identities obtained by successive integrations also reduces to the von K\'arm\'an momentum integral equation. 
The identities for free-stream boundary layers with a finite integration bound, \eqref{eq:fik-alternative-1}, are still useful for checking, numerically or experimentally, that the integrated $x$-momentum terms equate to the skin-friction coefficient computed via the wall-normal mean-velocity gradient at the wall. A further check is to verify that such equality holds irrespectively of the upper bound $h$, as long as the latter is located in the free stream, and of the number of integrations $n$, as we have shown for channel flows in figure \ref{fig:terms-3-4-5}(d).

For channel flows, only the original identity found by \cite{fukagata-iwamoto-kasagi-2002} possesses a physical meaning and we have proved that the infinite family degenerates to the definition of skin-friction coefficient as the number of integrations grows asymptotically. By a twofold integration, we have found an identity, valid for channel and pipe flows, that links the skin-friction coefficient with the integrated Reynolds stresses and the centreline mean velocity (the identity for channel flows was also discovered by \cite{elnahhas-johnson-2022}).

In the formula of the momentum thickness written as the sum of an energy thickness and an inertia thickness, we have expressed the latter as the sum of a thickness related to the mean-flow wall-normal convection and a thickness linked to the mean-flow streamwise inhomogeneity. This decomposition has been useful to further interpret the skin-friction decomposition of \cite{renard2016theoretical} physically and for quantifying the role of the different momentum-equation terms on the friction drag.

\section*{Acknowledgments}
\begin{acknowledgments}
We would like to thank the reviewers, Mr Ludovico Fossa', Dr Elena Marensi, Prof. Beverley McKeon, Mr Ahmed M.A. Elnahhas and Prof. Perry Johnson for the useful comments and the encouragement. PR has been partially supported by EPSRC (Grant No. EP/T01167X/1).
\end{acknowledgments}

\section*{Declaration of interests}
The authors report no conflict of interest.

\appendix
\section{Mean wall-normal velocity in the free stream}
\label{app:a}

In the derivation of equation \eqref{eq:fik}, the mean wall-normal velocity vanishes in the free stream, i.e. $\overline{v}$$\rightarrow$$0$ as $y$$\rightarrow$$\infty$. The von K\'arm\'an momentum integral equation \eqref{eq:theta-2} is instead obtained in \cite{hinze-1975} by assuming that the wall-normal velocity approaches a constant value (refer to his equation (7-8) on page 594 derived from the continuity equation). \cite{hinze-1975}'s assumption refers, however, to the first-order wall-normal velocity in the free stream: a wall-normal pressure gradient exists in the free stream to allow $\overline{v}$$\rightarrow$$0$ as $y$$\rightarrow$$\infty$. This adjustment is analogous to the second-order outer expansion in the case of a laminar boundary layer, where the solution is given in terms of a streamfunction obtained by complex-variable theory, as discussed in \cite{vandyke-1975} on page 135. 

Nevertheless, either choice for $\overline{v}$ in the free stream leads to \eqref{eq:theta-2}. Integrating \eqref{eq:rans-x} along $y$ from 0 to $\infty$ without adopting the boundary-layer approximation and by assuming $\lim_{y\rightarrow \infty}$$\overline{v}$$=$$\overline{v}_\infty$$\neq$$0$ 
leads to
\begin{equation}
\label{app:1}
\frac{\p \overline{u}}{\p y}\biggm|_{y=0} + 
\frac{\mathrm{d}}{\mathrm{d}x} \int_0^\infty \overline{u u} \dy +
\overline{v}_\infty -
\frac{\mathrm{d}}{\mathrm{d}x} \int_0^\infty \frac{\p \overline{u}}{\p x} \dy 
=0.
\end{equation}
Using the continuity equation and assuming that $\p \overline{u'u'}/\p x \ll \p \overline{u} \ \overline{u}/\p x$, the second term and the third term in \eqref{app:1} merge and the fourth term is written using $\overline{v}_\infty$, as follows
\begin{equation}
\label{app:2}
\frac{\p \overline{u}}{\p y}\biggm|_{y=0} =
\frac{\mathrm{d}}{\mathrm{d}x} \int_0^\infty \overline{u}(1-\overline{u}) \dy -
\frac{\mathrm{d}\overline{v}_\infty}{\mathrm{d}x}.
\end{equation}
Although \cite{hinze-1975} assumed that $\overline{v}_\infty$$\neq$$0$, the last term in \eqref{app:2} can be neglected because it derives from $\p^2 \overline{u}/\p x^2$ in \eqref{eq:rans-x}, which is negligible if the boundary-layer approximation is adopted, as on page 589 in \cite{hinze-1975}. 
In our analysis, \eqref{app:1} simplifies because $\overline{v}_\infty$$=$$0$ and, although $\p^2 \overline{u}/\p x^2$ is not neglected, the fourth term is null because it is equal to the last term in \eqref{app:2}. Equation \eqref{app:1} reduces to \eqref{app:2} because the second term in \eqref{app:1} becomes the first term on the right-hand side of \eqref{app:2} as the null term $-(\mathrm{d}/\mathrm{d}x)\int_0^\infty \overline{u} \dy$ can be reintroduced in \eqref{app:1}.

\section{Uniform convergence of integrands in integral relation \eqref{eq:fik-channel-alternative}}
\label{app:b}

In order to take the limit of \eqref{eq:fik-channel-alternative-2} as $n $$\rightarrow$$\infty$, we prove that the limiting operation can be transferred inside the integrals by using the dominated convergence theorem \citep{zeidler-etal-2012,pryce-2014}. Since the integration interval is finite and both integrand functions are bounded in this interval, it is sufficient to prove that the integrands converge uniformly. We first define $f_n = (y_c-1)^{n-1} \overline{u_c' v_c'}$. It is found that $\lim_{n \rightarrow \infty} f_n = f = 0$ because $(y_c-1)^{n-1}$$\rightarrow$$0$ for every $y_c \neq 0$ and $\overline{u_c' v_c'}(y_c=0)=0$. It follows that $\| f_n - f \|_{\infty} = \sup_{y \in [0,1]} \vert f_n \vert$$\rightarrow$$0$ as $n$$\rightarrow $$\infty$ because $f_n$ does so for every $y_c$. The proof for the integrand $(y_c-1)^{n-2} \overline{u}_c$ is analogous as $(y_c-1)^{n-2} $$\rightarrow$$0$ for every $y_c$$\neq$$0$ and $\overline{u}_c(y_c=0)=0$.

\bibliography{pr}
\bibliographystyle{jfm}

\end{document}